\documentclass{article}

\usepackage{arxiv}
\usepackage[
    backend=biber,
    style=vancouver,
    maxcitenames=10,
    maxbibnames=10,
    backref=true,
]{biblatex}

\DeclareFieldFormat{journaltitle}{\mkbibemph{#1}}
\DeclareFieldFormat[article]{journaltitle}{\mkbibemph{#1}}

\DeclareFieldFormat{shortjournal}{\mkbibemph{#1}}

\AtEveryBibitem{
  \clearfield{url}
  \clearfield{urldate}
  \clearfield{urlday}
  \clearfield{urlmonth}
  \clearfield{urlyear}
}

\DeclareFieldFormat[book]{title}{\mkbibemph{#1}}

\DeclareFieldFormat[article,periodical]{volume}{\mkbibbold{#1}}
\usepackage{amsmath}
\usepackage{graphicx}
\usepackage{float}
\usepackage{multicol}
\usepackage{listings}
\usepackage{framed}
\usepackage{capt-of}
\usepackage{newfloat}
\usepackage{subcaption}
\usepackage{hyperref}  

\usepackage{xcolor}
\definecolor{keywordcolor}{RGB}{100,0,100}
\definecolor{commentcolor}{RGB}{50,100,50}

\DeclareFloatingEnvironment[fileext=frm,placement={!ht},name=Listing]{listing}
\title{Quantum Integration Software Suite for HPC-Quantum Computing Workflows}

\author{
    {\hspace{1mm}Mateusz Meller}\thanks{Corresponding author -- mateusz.meller@stfc.ac.uk} \\
	STFC, Hartree Centre \\
	Daresbury, United Kingdom \\
	\And
    {\hspace{1mm}Joseph C. R. Thacker} \\
	STFC, Scientific Computing\\
	Daresbury, United Kingdom \\
    \And
    {\hspace{1mm}Thomas W. Keal} \\
	STFC, Scientific Computing \\
	Daresbury, United Kingdom \\
    \And
    {\hspace{1mm}Vendel Szeremi} \\
	STFC, Hartree Centre \\
	Daresbury, United Kingdom \\
}
    
\renewcommand{\shorttitle}{\textit{QISS -- A quantum integration software suite}}

\hypersetup{
pdftitle={Quantum Integration Software Suite for HPC-Quantum Computing Workflows},
pdfsubject={quantum computing, high-performance computing},
pdfauthor={Mateusz Meller, Joseph C. R. Thacker, Thomas W. Keal, Vendel Szeremi},
pdfkeywords={quantum computing, compilers, numerical simulation, performance models},
}

\begin{document}
\maketitle

\begin{abstract}
	In this work, we focus on addressing challenges related to modelling the integration of conventional high-performance computing (HPC) and quantum computing (QC). We introduce the \emph{Quantum Assembly Compiler (QASC)} for OpenQASM3.0 files. The compiler produces an executable that leverages the high-performance simulator QuEST to simulate quantum programs. Additionally, we introduce a performance modelling tool, \emph{QACC-perf}, which can estimate the cost of integrating QC in classical HPC workflows, as well as the execution time of the quantum circuits. Finally, we  demonstrate the utility of these tools in the context of multiscale quantum chemistry workflows using the ChemShell computational chemistry environment.
\end{abstract}


\section{Introduction}
The field of quantum computing (QC) leverages the underlying physics of quantum systems to perform computation. Theoretical work shows that for some problems, quantum computers could provide scaling advantage, providing solutions to problems intractable for classical computers \cite{nielsen_quantum_2010}.

Recent advancements in quantum computing hardware (\cite{serrano_quantum_2022}, \cite{bluvstein_logical_2023}, \cite{bruzewicz_trapped-ion_2019}, \cite{arute_quantum_2019}) motivate exploration of the scientific applications that could be accelerated by quantum computers (\cite{alexeev_quantum-centric_2024}, \cite{bauer_quantum_2020}, \cite{di_meglio_quantum_2023}). The specialised nature of the quantum devices and the expected classical overhead of error-correction imply that quantum computers will need to work with classical High-Performance Computing (HPC) systems in tandem \cite{humble_quantum_2021-1}.

From the perspective of scientific software running on the supercomputers, the adoption of quantum computing is expected to be incremental. Because of the existing value and size of scientific code, rewriting rapidly those projects is often unrealistic. Instead, the most promising kernels within the code can be identified for porting and offloading to Quantum Processing Units (QPUs), as is currently the case with Graphical Processing Units (GPUs) \cite{buchs_role_2026}, \cite{johansson_quantum_2021}, \cite{perez_castillo_software_2021}. As such, there is a need to analyse existing workflows and evaluate the approaches for HPC-QC integration \cite{elsharkawy_integration_2025}.
Moreover, given the limited access and the cost of quantum computers, researchers use simulators and emulators for verification. The cost of simulation is high, however. Therefore, borrowing from the performance modelling methods used in classical HPC (\cite{williams_roofline_2009}, \cite{culler_logp_1996}, \cite{ino_loggps_2001}), it is reasonable to assume that similar methods are needed for quantum computing.

In this work, we focus on addressing those points through two new software tools. First, we present the Quantum Assembly Compiler (QASC) for OpenQASM3.0 \cite{cross_openqasm_2022} files. The main objectives for the design of the QASC were to create tool that is scalable, portable and interoperable with major quantum computing frameworks. The compiler produces an executable that leverages the high-performance simulator QuEST \cite{jones_quest_2019} to simulate quantum programs. Additionally, we introduce a performance modelling tool, QACC-perf, which can estimate the cost of integrating QC in classical HPC workflows, as well as the execution time of quantum circuits. To summarise, this paper makes the following contributions:

\begin{enumerate}
    \item A source‑to‑source compiler enabling efficient simulation of general OpenQASM3.0 programs with classical control.
    \item A lightweight performance model tool for evaluating HPC–QC offloading strategies.
    \item Benchmarking of the tools in the context of a realistic, production‑scale workflow for computational chemistry.
\end{enumerate}

\section{Related Work}
At the lowest level, quantum circuits are described by quantum assembly languages which are hardware-agnostic. One of the most prominent examples is the OpenQASM family of languages developed by IBM \cite{cross_openqasm_2022}. While OpenQASM2.0 \cite{cross_open_2017} could describe any quantum computation, a broader set of primitives was required for tasks such as error-correction. OpenQASM3.0 \cite{cross_openqasm_2022} has been introduced as an extension to OpenQASM2.0, which supports an extensive set of classical operations. These include loops, branching and input/output. This allows for mid-circuit measurement branching with complex logic, enabling the specification of a new class of algorithms. However, to date, there has been a limited number of compilers or simulators that accept OpenQASM3.0. Notably, IBM introduced the “qasm3” module into their quantum computing framework, Qiskit \cite{aleksandrowicz_qiskit_2019}. The module allows for circuit definition, manipulation, simulation and transpilation. Although still a work in progress, the module is integrated with the rest of the framework.

Recently, Pennylane \cite{bergholm_pennylane_2022} incorporated functionality into their framework, allowing parsing of the OpenQASM3.0 circuits and interpreting them. Compared to their contribution, here we focus on compiling the circuits for the further scalable simulation. 
An alternative simulator, which accepts a subset of OpenQASM3.0, including
classical branching (for a mid-circuit measurement), is DDSIM, part of a Munich Quantum Toolkit, developed under the umbrella project Munich Quantum Software Stack (MQSS) \cite{kaya_software_2024}. MQSS itself is a project spanning compilation, simulation and resource estimation.

Compared to the existing work, we propose lightweight, plug-and-play tooling for compilation and latency estimation, which can be easily deployed and is agnostic to the parent project’s choice of programming language.

\section{QASC: OpenQASM3-to-QuEST Compilation}
    The \emph{Quantum Assembly Compiler (QASC)} is a source-to-source compiler targeting
    the OpenQASM3.0 specification. The motivation for developing it comes from two
    pain points common in the development and execution of quantum algorithms.
    First, OpenQASM2.0 serves as a \emph{de facto} standard low-level representation of
    quantum circuits. In the era of noisy intermediate-scale quantum (NISQ) devices
    \cite{preskill_quantum_2018} and shallow circuits,
    it serves its purpose. However, when exploring fault-tolerant algorithms
    of the future (for example, quantum phase estimation for chemistry \cite{bauer_quantum_2020}),
    the circuits (and correspondingly, the files) grow to prohibitive sizes. 
    The bottleneck is the lack of classical constructs to represent loops
    and branching. Moreover, because of this, OpenQASM2.0 cannot be used for
    circuits that contain mid-circuit measurement and complex logic.
    OpenQASM3.0 was introduced to alleviate these issues, and indeed, the
    specification is a complete language, supporting ``for loops'', ``branching'',
    quantum gates, and even gate calibration.

    Although OpenQASM3.0 can be generated from the major quantum computing frameworks, with this power comes complexity. Because of the scope of OpenQASM3.0, there is currently less vendor support for processing it. As such, aside from the
    Qiskit experimental implementation described above, there are no open-source tools for compiling it and efficiently simulating the resulting circuits, creating a barrier for
    the exploration of fault-tolerant circuits. \emph{QASC} addresses those issues
    by compiling OpenQASM3.0 files into C++ files, simulating quantum gates with the
    QuEST routines. Crucially, \emph{QASC} is a lightweight, self-contained
    tool that can easily be installed on supercomputers and used for simulation
    of arbitrary quantum programs.

    We have chosen to target QuEST as it is a high-performance library written in C++, used for simulating
    quantum state-vectors and density matrices. It supports multithreading,
    distributed computing and GPU acceleration via a combination of OpenMP, MPI,
    CUDA, HIP, Thrust, and cuQuantum. Thanks to its scalability
    and performance characteristics, it is suitable for simulating even the
    most demanding quantum circuits.

    \subsection{QASC Architecture}

\begin{figure}[ht!]
        \begin{center}
            \includegraphics[width=0.95\textwidth]{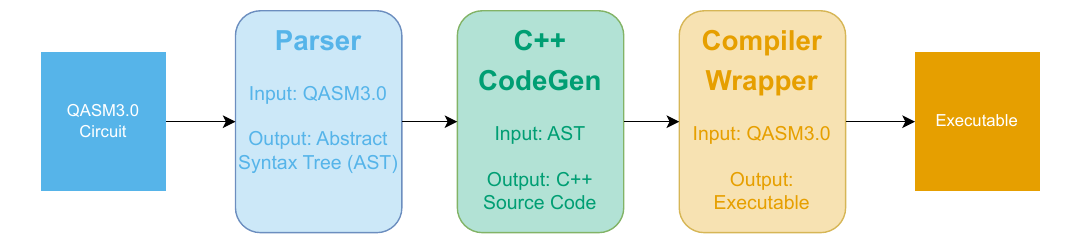}
        \end{center}
        \caption{Diagram showing the compilation process using \emph{QASC}.
                 The OpenQASM3.0 source is processed by the parser which produces an
                 abstract syntax tree (AST). The next component generates
                 C++ code from the AST. Then the code is compiled with a C++ compiler
                 (e.g. clang++ or g++) and finally linked with QuEST library. }
        \label{fig:qasc-compiler-flow}
    \end{figure}

    The compiler is built of three main components: -- a parser, a C++ code generator
    and a compiler wrapper, which compiles the generated code under the hood.
    Figure \ref{fig:qasc-compiler-flow} shows the schematic of the 
    compilation process.

    \subsubsection{Parser}
    For the parser, we leverage the Antlr4 \cite{parr_definitive_2013} project to generate corresponding
    Lexer and Parser components according to the grammar and tokens outlined 
    in the OpenQASM3.0 specification. Even though from the performance point of view
    having a dedicated parser would be desirable, a) the parsing is not a
    bottleneck, b) Antlr4 is an established framework with a lot of flexibility
    that makes it easier to extend this work and add optimisation passes.
    Under the hood, Antlr4 builds an abstract syntax tree (AST) and provides 
    visitors for traversing the tree, allowing operations on the nodes. For
    our purpose, the operation applied to the node is C++ code generation.

    \subsubsection{C++ Code Generation}
    Having built the AST, the next step is the C++ Code Generator. It is a
    specialised visitor class inheriting from the ParserVisitor generated by
    Antlr4. For the classical statements (e.g. ``if-else'', function definitions, loops),
    it performs translation ensuring that the resulting code is a valid
    C++ code. The classical arrays are represented as standard library vectors
    storing elements of the corresponding type.

    For the quantum constructs, we redefine the OpenQASM3.0 qubit data type to be 
    QuEST's Qureg data structure.  Additionally, to use QuEST, one needs to initialise its environment (which stores information about the workers and modes of parallelisation). The compiler generates the required environment
    setup code on the first contact with the quantum declaration. Internally, this
    sets a flag, ensuring that the initialisation step is generated only once
    per program.
    To rewrite calls to quantum gates, the generator stores a dictionary
    with keys being quantum gates defined by OpenQASM3.0 specification, and
    values are corresponding QuEST gate routine names, along with the
    information about the type and number of function arguments.
    Based on the quantum gate, the generator correctly deduces the arguments
    and rewrites them into a QuEST routine call.

    Output statements are implemented, which convey the intention
    that the associated variable should be returned at the end of the quantum
    program. Here, it is handled by recording the name of the variable and generating
    code which writes the contents of the variable into the output file.
    Likewise, input statements are implemented by tracking the input variable and
    populating it with the contents of the file, provided as a command-line argument.
    Optionally, at the end of the code generation process, the code can
    be formatted if there is \emph{clang-format} in the path. This feature is
    controlled by a command-line flag.

    \subsubsection{Intermediate C++ Compilation}
    The example C++ code resulting from the code generation step can be seen
    in Listing \ref{lst:cppqft}.
    The final stage of the compilation is the actual compilation of the resulting
    C++ source file. This is achieved by calling a C++ compiler, such as \emph{g++}
    or \emph{clang++} and then linking to the QuEST library.
    The user can either build the project and specify the desired C++ compiler
    at compile time or pass the path to the compiler via a command-line argument.
    Similarly, the user should pass the path to the QuEST library during the
    project compilation stage or via the command-line.

    \subsubsection{Usage}
    Below, we outline the usage of \emph{QASC} from the command line.
    Table \ref{tab:qasc-usage} details the available parameters, along
    with the function of each flag.

    \begin{verbatim}
    Usage: qasc [--help] [--version] -cxx VAR --input VAR
                --output VAR [-L VAR] [-rpath VAR] [-I VAR]
                [-F] [--verbose]
    \end{verbatim}

    \begin{table}[ht!]
        \begin{center}
            \begin{tabular}[c]{l|l}
                \hline
                \multicolumn{1}{c|}{\textbf{Command-line Flag}} & 
                \multicolumn{1}{c}{\textbf{Help Message}} \\
                \hline
                -h, --help    & shows help message and exits \\ 
                -v, --version & prints version information and exits \\ 
                -cxx          & specify standard C++ compiler. [required] \\ 
                -i, --input   & specify the input file. [required] \\ 
                -o, --output  & specify the output file. [required] \\ 
                -L            & specify path to quest library. [nargs=0..1] [default: ""] \\ 
                -rpath        & specify rpath to quest library. [nargs=0..1] [default: ""] \\ 
                -I            & specify path to quest include. [nargs=0..1] [default: ""] \\
                -F            & set output cpp formating (requires clang-format). \\
                -V, --verbose & set qasc program execution to verbose mode. \\
                \hline
            \end{tabular}
        \end{center}
        \caption{Command-line options for the \emph{QASC}.}
        \label{tab:qasc-usage}
    \end{table}

    \subsection{Language Features}

    For the complete description of the OpenQASM3.0 constructs, we refer the reader
    to the article introducing OpenQASM3.0 \cite{cross_openqasm_2022} and the associated
    website\footnote{https://openqasm.com/}. Our solution
    supports most of the constructs outlined there, except a few, described below.

    \subsubsection{Limitations}

    Currently \emph{QASC} does not support calibration statements or physical qubits. These features of OpenQASM3.0 are required for complex experiments, enabling precise and efficient control of the underlying quantum hardware. 

    The decision not to support those features was dictated by our internal use cases, the fact that our workflows focus on general quantum circuits rather than the emulation of specific hardware, and the fact that simulating them in QuEST would be inefficient. The reason is that the gates are calibrated with pulses and target physical qubits. Simulating pulses can be challenging from a performance point of view. Future work will extend \emph{QASC} to support these features.

    Additionally, the current implementation does not support ``pragmas'' and ``annotations'' as defined in OpenQASM3.0 specification. These are not required to run valid circuits and are considered an extension to \emph{QASC}, left for future work.

    \subsection{Performance Characteristics}
    From the performance point of view, the compiler relies heavily on the parser implementation from Antlr4 and the underlying C++ compiler. Table \ref{tab:results-compilation} shows the timings of the different components of the toolchain. The memory usage is directly associated with the size of the OpenQASM3.0 file and the memory overheads of the tools. Internally, the generated code is stored in a dynamic string buffer.
    When it comes to simulation, QuEST is a high-performance framework which is efficient and scalable. For the exact details about the implementation of QuEST the reader is referred to \cite{jones_quest_2019}.
    \newpage

            \begin{lstlisting}[gobble=20, language=C, label={lst:qasm3qft}, escapechar=|, caption={OpenQASM3.0 implementation of QFT.}, captionpos=b]
                int num_shots = 10;
                int num_qubits = 4;
                    qubit[num_qubits] qubits;
                    output array[int, num_shots, num_qubits] measurements;

                for int shot in [0 : num_shots] {
                    reset qubits;
                    for int i in [0 : num_qubits] {
                        h qubits[i];

                        float param = pi / 2;
                        for int j in [i + 1 : num_qubits] {
                            crz(param) qubits[j], qubits[i];
                            param /= 2;
                        }

                    }

                    for int i in [0 : num_qubits / 2] {
                        int j = num_qubits - (i + 1);
                        swap qubits[i], qubits[j];
                    }

                    for int i in [0 : num_qubits] {
                         measurements[shot][i] = measure qubits[i];
                    }
                }
            \end{lstlisting}

            \newpage
                \begin{lstlisting}[gobble=20, language=C, label={lst:cppqft}, escapechar=|, caption={\emph{QASC} generated C++ code from Listing 1.}, captionpos=b]
                    #include <algorithm>
                    #include <cmath>
                    #include <fstream>
                    #include <iostream>
                    #include <type_traits>
                    #include <vector>
                    #include "argparse.hpp"
                    #include "qasc_gates.h"
                    #include "quest.h"

                    using namespace qasc;
                    QasmType<int, 32> num_shots(10);
                    QasmType<int, 32> num_qubits(4);
                    int main(int argc, char* argv[]) {
                        auto qubits = QuregSlice(num_qubits);
                        if (!isQuESTEnvInit()) {
                            initQuESTEnv();
                        }
                        if (!isQascInternalQuestQuregInit) {
                            isQascInternalQuestQuregInit = true;
                            qascInternalQuestQureg = createQureg(0 + num_qubits);
                        }
                        std::vector<std::vector<QasmType<int, 32>>> measurements(
                            num_shots, std::vector<QasmType<int, 32>>(num_qubits, 0));
                        for (auto shot = 0; shot < num_shots; shot += 1) {
                            reset(qubits);
                            for (auto i = 0; i < num_qubits; i += 1) {
                                h(qubits[i]);
                                double param = std::numbers::pi / 2;
                                for (auto j = i + 1; j < num_qubits + 1; j += 1) {
                                    crz(param, qubits[j], qubits[i]);
                                    param /= 2;
                                }
                            }
                            for (auto i = 0; i < num_qubits / 2; i += 1) {
                                QasmType<int, 32> j(num_qubits - (i + 1));
                                swap(qubits[i], qubits[j]);
                            }
                            for (auto i = 0; i < num_qubits; i += 1) {
                                measurements[shot][i] = one_qubit_measure(qubits[i]);
                            }
                        }
                        std::ofstream out("qft.results");
                        out << "measurements" << ":\n";
                        for (auto& a : measurements) {
                            for (auto& b : a) {
                                out << b << ",";
                            }
                            out << "\n";
                        }
                        out << "\n";
                        out.close();
                        finalizeQuESTEnv();

                        return 0;
                    }
                \end{lstlisting}

    \subsection{Examples}
    To demonstrate \emph{QASC} in action, in the project’s repository \cite{meller_qasc_2026}, we have included a set of examples implementing different quantum algorithms. As an example, above we show the input OpenQASM3.0 file implementing the Quantum Fourier Transform (QFT) and the resulting generated C++ code. Further validation of the proposed solution was achieved by the mix of unit tests and known quantum circuits.

\section{QACC-Perf: Latency and Runtime Estimation}
The second software tool we present is a performance modelling toolkit called \emph{QACC-Perf}. The main application of this tool is to explore different HPC-QC integration modes. As such, it can guide the development of hybrid quantum-classical algorithms and provide insight into resource constraints. This is achieved by estimating the runtime, communication overheads and system latency for a given quantum circuit. Moreover, in the context of our case study, it can serve as a plug-and-play component of a wider quantum chemistry workflow, which calculates molecules' ground state energy and waits for the estimated time to simulate the offloading procedure.

    \subsection{Modelling Approach}

    The model for estimating the cost of the offloading is made of three major
    components: the latency of the network based on the HPC-QC integration type,
    the bandwidth of the used network and the runtime on the actual quantum
    device. The general formula for the kernel offloading time is described as:
    \begin{equation}
        K = l + T + R 
        \label{eq:perfmodel}
    \end{equation}
    where $l$ is the total communication latency,
    $T$ is the data transfer cost, and $R$ is the total quantum runtime.
    The derivation of each of those parameters is provided in more detail below.

    \subsection{Application Configuration}
    QACC-Perf expects an input quantum circuit, a JSON-based
    configuration file and path to the log file which stores the timings.
    Optionally, it takes a flag which specifies if the circuit is in OpenQASM3.0
    format. Another flag specifies if the tool should make use of an OSU-based \cite{breuer_jupiter_2024} estimator,
    which calculates the latency and bandwidth of the network.
    The format of the JSON configuration file is outlined in Table
    \ref{tab:qaccjsonformat}.

    \begin{table}[t!]
        \begin{center}
            \begin{tabular}[c]{l|l}
                \hline
                \multicolumn{1}{c|}{\textbf{Key}} & 
                \multicolumn{1}{c}{\textbf{Values}} \\
                \hline
                integration & str: \{cloud, multi-node, on-node\} \\
                network     & str: \{cloud, ethernet, infiniband\} \\
                arch        & str: \{star, grid, all-to-all, hex, custom\} \\
                gates       & [\{name: gate name, duration: float, error: float\}, ...] \\
                gates       & [\{name: gate name, connectivity: (q0: int, q1: int), \\
                (if arch is custom) & duration: float, error: float\}, ...]\\
                shots       & int \\
                workers     & int \\
                fcidump     & str \\
                
                \hline
            \end{tabular}
        \end{center}
        \caption{Format of the JSON configuration file used by QACC-Perf.}
        \label{tab:qaccjsonformat}
    \end{table}

    \subsection{Latency}
    
    The latency component can be derived in two ways.
    First, the user can provide the system specification and allow the tool to measure the latency by performing a ping to the destination.
    This is implemented via \emph{OSU microbenchmarks} \cite{breuer_jupiter_2024}.
    Alternatively, we provide default values for the ``cloud'' integration,
    ``multi-node'' latency in a local system and ``on-node'' communication.
    The values are based on running and averaging the OSU microbenchmarks on a HPC cluster comprised of dual AMD EPYC 9825 144-core 2.2 GHz processors and HPE Slingshot 11 interconnects. The values are outlined in Table \ref{tab:hardcodedvals}.
    \begin{table}[t!]
        \begin{center}
            \begin{tabular}[c]{l|r|r}
                \hline
                \multicolumn{1}{c|}{\textbf{Integration type}} & 
                \multicolumn{1}{c|}{\textbf{Latency [s]}} & 
                \multicolumn{1}{c}{\textbf{Bandwidth [MB/s]}} \\
                \hline
                on-node     & 1.5e-5 & 68670   \\
                multi-node  & 4e-5   & 23375    \\
                cloud       & 3e-2   & 50      \\
                
                \hline
            \end{tabular}
        \caption{Default values used for the communication overhead depending on the integration mode.}
        \label{tab:hardcodedvals}
        \end{center}
    \end{table}
    Then the total latency $l$ is given by multiplying the estimated latency
    by two (in order to account for sending and receiving operations).
    This is an approximation of real world systems and enables \emph{QACC-Perf} to give a general understanding of the qualitative overheads associated with the integration.

    \subsection{Data Transfer}

    The overhead of data transfer is computed first by estimating the
    network bandwidth $B$ and then calculating the message size $S$. Here,
    the message size is the size of the circuit.
    This is an idealised scenario to give a qualitative estimate.
    Similarly to latency, the bandwidth $B$ can be estimated in two ways.
    Either by running a script based on the OSU implementation,
    which measures the network bandwidth using a data transfer,
    or by using pre-defined values.
    For the default values the available options are ``cloud'', ``multi-node'' and ``on-node''. The values were obtained similarly to latency values and are outlined in Table \ref{tab:hardcodedvals}. Then we have $T=S/B$.

    \subsection{Runtime Prediction}

    For the runtime prediction, the tool depends on  Qiskit library \cite{aleksandrowicz_qiskit_2019}, which allows for estimating the duration of the transpiled circuit. This is achieved by building a directed-acyclic graph of the circuit and measuring longest weighted path, where, weight of the edge is duration of the gate. The computed results are based on the configuration file supplied by the user and the gate durations within. The input circuit is first transpiled to the expected topology and basis gate set. Then the execution time of the circuit is estimated. Moreover, the transpilation step is timed itself and given at the end of the modelling.
    Given the circuit execution time $t$, number of shots $s$ and
    the number of quantum workers $W$, the total runtime is given by
    $R = (s * t) / W$.

    \subsection{Limitations}
    
    Even though the QACC-Perf has been successfully used in a project, it has
    limitations. The first one is due to the incomplete implementation of the
    ``qasm3'' module in Qiskit, at the time of writing. As a result, estimating
    the offloading cost of general quantum circuits is limited to OpenQASM2.0 files.
    While it does support OpenQASM3.0 files, it is mostly tailored towards
    one specific use case and should be considered experimental.
    Another limitation is that the default values used to estimate
    latency and bandwidth are idealised based on the vendor's specification.
    In the real world, there are multiple factors that affect both the latency
    of the system and the available bandwidth for data transfer. As such,
    results obtained from this tool should be used to get a general,
    high-level understanding rather than concrete numbers.

\section{Case Study: ChemShell Integration}
The tools presented in this work have been used in the extension of the ChemShell chemistry package \cite{lu_open-source_2019} to support hybrid HPC-QC computation. Here, we will cover only the details of the framework that relate to the new tools introduced in this paper. For the broader multiscale embedding framework, we refer the reader to our companion paper \cite{thacker_hybrid_nodate}. We note that the \emph{QASC} compiler presented here has been extended and generalised based on this initial use case.

    \subsection{ChemShell Overview}

    ChemShell is a computational quantum chemistry package which allows users to model complex chemical systems, focussing on hybrid quantum mechanics/molecular mechanics (QM/MM) embedding. The implemented computational methods are highly performant, efficient and scalable. Due to its modular design, it can easily be interfaced with existing quantum and classical software, enabling a mix-and-match approach for combining quantum calculations with fast molecular mechanical simulations across a range of software. As such, it serves as a computation orchestrator, handling work distribution, communication and the detailed bookkeeping required to combine diverse calculation types.
    In our companion paper \cite{thacker_hybrid_nodate}, we extended ChemShell with a multiscale embedding framework that combines QM/MM calculations on conventional HPC platforms with high level quantum computing algorithms incorporating orbital active space embedding. The \emph{QASC} and \emph{QACC-Perf} tools introduced in this work perform a crucial role within the new framework, as we describe here.

    \subsection{Integration Scenario}

    In this case study, \emph{QASC} was introduced as a part of an effort to
    combine the Quantum Phase Estimation (QPE) algorithm within the multiscale
    embedding approach.

\begin{figure}[t!]
        \begin{center}
            \includegraphics[width=0.45\textwidth]{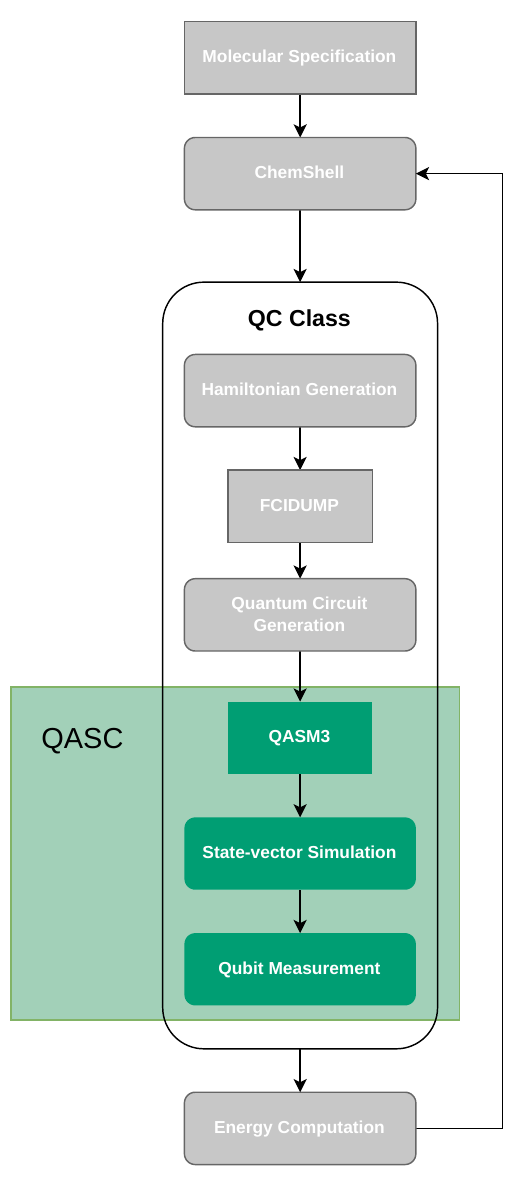}
        \end{center}
        \caption{Outline of the workflow implemented in ChemShell as presented in our companion article \cite{thacker_hybrid_nodate}.
                The highlighted part is enabled by the work introduced in this paper.
                First, the
                 users provide molecule specifications, such as geometry, to ChemShell.
                 Then the information is passed into the ``QC class'', which generates the Hamiltonian
                 and associated FCIDUMP file. Next, the quantum circuit is generated,
                 resulting in a OpenQASM3 file. The OpenQASM circuit is then compiled, and the
                 execution is simulated with the QuEST simulator. Finally, qubits are
                 measured, and from the histogram of measurements the system calculates the energy,
                 which is passed back to ChemShell for another iteration step.
             }
        \label{fig:chemshell-workflow}
    \end{figure}

    The overall workflow is summarised in Figure \ref{fig:chemshell-workflow}. As the overall workflow is not the focus of the current work, we only describe it briefly here, and the interested reader should consult Ref. \cite{thacker_hybrid_nodate} for further details. First, a molecular specification is passed to ChemShell, where parameters such as geometry, charge and basis set are defined. Then, ChemShell passes the configuration to a quantum chemistry package, in this case PySCF \cite{sun_recent_2020}, which runs a Hartree-Fock calculation. The result of this step is an FCIDUMP file, which describes the terms of the electronic Hamiltonian. From this file, the next stage of the workflow generates a quantum circuit in OpenQASM3.0 format which implements iterative QPE according to the Somma et al. method \cite{somma_quantum_2019}.

    At that stage, the workflow calls \emph{QASC} and compiles the OpenQASM3.0 circuit.
    The compiled executable performs a distributed state-vector simulation.
    Depending on the available compute resources, it can employ additional
    shot-level parallelism.

    Next, the collected measurements from the simulation are interpreted to produce the resulting ground state energy, which is then returned to ChemShell for another iteration of geometry optimisation.
    Alternatively, instead of running a state-vector simulation, the user can
    plug in the \emph{QACC-Perf} tool, which reads in a generated OpenQASM3.0 circuit, estimates
    the execution time and overheads, then sleeps for this time, and returns the 
    ground state energy (computed via classical method). In this way, \emph{QACC-Perf}
    emulates offload to the quantum device and can be used in the benchmarking
    of the workflow as if it was using a real physical QPU.

    \subsection{Benchmarking Setup}

    To assess the performance of \emph{QASC}, we have recorded compilation timings
    for five different quantum circuits,
    of varying size, each performing iterative QPE to calculate the ground state
    energy of a different molecule. The selected ``backend'' for the compilation
    was LLVM clang++, version 20.1.8. The compilation was performed on a 
    2.3 GHz Quad-Core Intel Core i7 with 16 GB 3733 MHz LPDDR4X memory.
    The experiments were repeated five times, and the average has been presented.
    The runs were executed on a distributed system with each node consisting of 32 Intel(R) Xeon(R) Gold 6142 CPU at 2.60 GHz.
    The runtime estimation was based on the bandwidth and latency data from
    a different distributed system with Ethernet network MT2894 ConnectX-6 LX
    by Mellanox Technologies with 25Gbit/s capacity, width of 64 bits and
    clock at 33 MHz. The quantum gate basis set and the gate timings were taken
    from Ref. \cite{jordan_quantum_2026} and are outlined in Table \ref{tab:seeqc-gates}.

    \begin{table}[t!]
        \begin{center}
            \begin{tabular}{l|r|r}
                \hline
                \multicolumn{1}{c|}{\textbf{Gate Name}} &
                \multicolumn{1}{c|}{\textbf{Duration [ns]}} &
                \multicolumn{1}{c}{\textbf{Error Rate [\%]}} \\
                \hline
                $R_x$       & 51.8  & 0.272     \\
                $R_y$       & 51.8  & 0.272     \\
                $CZ$        & 82.0    & 1.625     \\
                $Measure$   & 4550.0  & 0.380      \\
                \hline
            \end{tabular}
        \end{center}
        \caption{Basis gate set and quantum gates details based on the \cite{jordan_quantum_2026}.}
        \label{tab:seeqc-gates}
    \end{table}

    \subsection{Results}
    In this section, we have compiled results for the compilation of the different circuit sizes, where by circuit size we mean the size of the OpenQASM3.0 file, which is proportional to the depth of the circuit. We have also included simulation timings along with the offloading times estimated by \emph{QACC-Perf}. 

    \begin{table}[t!]
        \begin{center}
            \begin{tabular}[c]{l|r|r|r|r|r}
                \hline
                \multicolumn{1}{c|}{\textbf{Molecule}} &
                \multicolumn{1}{c|}{\textbf{Circuit}} &
                \multicolumn{3}{c|}{\textbf{Compilation Time [s]}} &
                \multicolumn{1}{c}{\textbf{Memory}} \\
                \cline{3-5}
                
                \multicolumn{1}{c|}{\textbf{}} & 
                \multicolumn{1}{c|}{\textbf{Size [MB]}} & 
                \multicolumn{1}{c|}{\textbf{Frontend}} & 
                \multicolumn{1}{c|}{\textbf{Backend}} &
                \multicolumn{1}{c|}{\textbf{Total}} &
                \multicolumn{1}{c}{\textbf{Usage [MB]}} \\
                \hline
                $LiH$  &  0.6  & 2190.9 & 9198.69  &  11.45  & 345.466  \\
                $NH_3$ &  4.5  & 15.862 & 427.280  & 443.24  & 2033.995 \\
                $CH_4$ &  11.1 & 37.364 & 2584.402 & 2621.96 & 4530.820 \\

                \hline
            \end{tabular}
        \end{center}
        \caption{Results of the compilation of different quantum circuits
                implementing ground state energy estimation for a given molecule.
                The second column shows size of the circuit file corresponding to the QPE implementation for a given molecule.
                Columns 3-5 show compilation times of the compiler frontend,
                backend and total, respectively. Rightmost column shows memory
                usage as given by GNU `time'.}
        \label{tab:results-compilation}
    \end{table}
    
    \subsubsection{QASC Performance}
    Table \ref{tab:results-compilation} shows the averaged results of the compilation.
    The compilation time is directly proportional to the circuit size, as is
    the memory consumption. Even for the largest circuits, the overhead of the frontend
    is minimal, whereas the bulk of the computation is taken by the backend. The long
    computation can be attributed to the size of the resulting source files, and indeed,
    the initial quantum circuit. For those, the size comes predominantly from the unrolled
    controlled unitary, which implements Trotterized time-evolution under the molecular Hamiltonian.
    The presented memory consumption values are also predominantly associated with
    the backend compilation step.

    \begin{table}[t!]
        \begin{center}
            \begin{tabular}[c]{l|r|r|r|r}
                \hline
                \multicolumn{1}{c|}{\textbf{Molecule}} & 
                \multicolumn{1}{c|}{\textbf{Circuit}} & 
                \multicolumn{1}{c|}{\textbf{Compilation}} & 
                \multicolumn{1}{c|}{\textbf{Circuit}} & 
                \multicolumn{1}{c}{\textbf{Total}} \\

                \multicolumn{1}{c|}{\textbf{}} & 
                \multicolumn{1}{c|}{\textbf{Generation [s]}} & 
                \multicolumn{1}{c|}{\textbf{Time [s]}} & 
                \multicolumn{1}{c|}{\textbf{Simulation [s]}} & 
                \multicolumn{1}{c}{\textbf{Time [s]}} \\

                \hline
                $LiH$  & 13     &  10   & 605   & 628   \\
                $NH_3$ & 263    &  438  & 40213 & 40914 \\
                \hline
            \end{tabular}
        \end{center}
        \caption{Timings of a full ChemShell workflow per shot, per iteration.
                The workflow is split into steps, which include, circuit
                generation, compilation and circuit simulation. Last column shows the total time of all steps.}
        \label{tab:results-mol-timings}
    \end{table}

    For context, Table \ref{tab:results-mol-timings} outlines the timings of the different steps in
    a general ChemShell workflow with respect to the studied molecule. Due to exponential
    scaling with the number of qubits, the state vector simulation dominates
    the compute time. For a more detailed discussion of the full workflow, we refer the reader to the companion work \cite{thacker_hybrid_nodate}.

    \begin{table}[t!]
        \begin{center}
            \begin{tabular}[c]{l|r|r|r}
                \hline
                \multicolumn{1}{c|}{\textbf{Molecule}} & 
                \multicolumn{1}{c|}{\textbf{Circuit Size [MB]}} &
                \multicolumn{1}{c|}{\textbf{Latency [$\mu s$]}} &
                \multicolumn{1}{c}{\textbf{Bandwidth [MB/s]}} \\
                \hline
                $LiH$        & 0.6 &   25.43   &  18062.20   \\
                $NH_3$       & 4.5   &   384.06  &  10060.74  \\
                $CH_4$       & 11.1  &   2390.28 &  2411.04   \\
                $H_2O$ dimer & 50.3  &   9984.82 &  4430.38   \\
                \hline

            \end{tabular}
        \end{center}
        \caption{Estimated latency and bandwidth for the intra-node (on-node option selected) integration based on the quantum circuit size.}
        \label{tab:results-qacc-on-node}
    \end{table}

    \begin{table}[t!]
        \begin{center}
            \begin{tabular}[c]{l|r|r|r}
                \hline
                \multicolumn{1}{c|}{\textbf{Molecule}} & 
                \multicolumn{1}{c|}{\textbf{Circuit Size [MB]}} &
                \multicolumn{1}{c|}{\textbf{Latency [$\mu s$]}} &
                \multicolumn{1}{c}{\textbf{Bandwidth [MB/s]}} \\
                \hline
                $LiH$         & 0.6 &   325.63   & 2592.88   \\
                $NH_3$        & 4.5   &   1673.23  & 2478.89   \\
                $CH_4$        & 11.1  &   4486.39  & 2906.08   \\
                $H_2O$ dimer  & 50.3  &   19628.43 & 2867.95   \\
                \hline

            \end{tabular}
        \end{center}
        \caption{Estimated latency and bandwidth for the inter-node (multi-node option selected) integration based on the quantum circuit size.}
                \label{tab:results-qacc-multi-node}
    \end{table}

    \begin{table}[t!]
        \begin{center}
            \begin{tabular}[c]{r|r}
                \hline
                \multicolumn{1}{c|}{\textbf{Latency [$\mu s$]}} &
                \multicolumn{1}{c}{\textbf{Bandwidth [MB/s]}} \\
                \hline
                  79000  & 18.75   \\
                \hline

            \end{tabular}
        \end{center}
        \caption{Estimated latency and bandwidth for the cloud integration
                The overhead is dominated by the networking protocols and
                remains constant for the studied circuit sizes. }
                \label{tab:results-qacc-cloud}
    \end{table}

    Table \ref{tab:results-qacc-on-node}, Table \ref{tab:results-qacc-multi-node} and Table \ref{tab:results-qacc-cloud}
    show the results from \emph{QACC-Perf}, which estimates latency and bandwidth
    depending on the QPU-HPC integration mode (intra-node/on-node, inter-node/multi-node, cloud) and the circuit size.

    For the intra-node integration (Table \ref{tab:results-qacc-on-node}),
    while the latency is minimal and the bandwidth is the largest, it
    progressively saturates with the size of the circuit. At some point,
    the overhead starts to be comparable with the inter-node communication.

    Regarding the inter-node integration (Table \ref{tab:results-qacc-multi-node}), 
    while the latency increases with the size of the quantum circuit,
    the bandwidth remains relatively constant. Compared to the intra-node
    communication, the ratio between latencies falls from 12.8x slowdown
    to roughly 2x slowdown, which implies that the performance hit is the
    largest for the small circuits.

    One can see that for the cloud integration (Table \ref{tab:results-qacc-cloud}),
    while the latency is roughly between one (inter-node integration)
    and two (intra-node integration) orders of magnitude larger, both the
    latency and bandwidth are assumed to be constant with respect to the circuit size.
    The overheads associated with the internet connection, networking protocols and
    routing dominate here, and for the circuit sizes we consider, the network is
    already under heavy load.

    \begin{table}[t!]
        \begin{center}
            \begin{tabular}[c]{l|r|r|r}
                \hline
                \multicolumn{1}{c|}{\textbf{Molecule}} & 
                \multicolumn{1}{c|}{\textbf{On-node [ms]}} &
                \multicolumn{1}{c|}{\textbf{Multi-node [ms]}} &
                \multicolumn{1}{c|}{\textbf{Cloud [ms]}} \\
                \hline
                $LiH$        &  0.084  & 0.880  & 189.573  \\
                $NH_3$       &  1.206  & 5.122  & 392.667  \\
                $CH_4$       &  9.343  & 12.758 & 744.667  \\
                $H_2O$ dimer &  30.804 & 55.994 & 2718.000 \\
                \hline
            \end{tabular}
        \end{center}
        \caption{Comparison of the communication overhead based on the
                integration mode with respect to the quantum circuit size.}
        \label{tab:results-qacc-comms}
    \end{table}

    \begin{figure}[t!]
    \centering
        \begin{subfigure}{.45\textwidth}
          \centering
          \includegraphics[width=\linewidth]{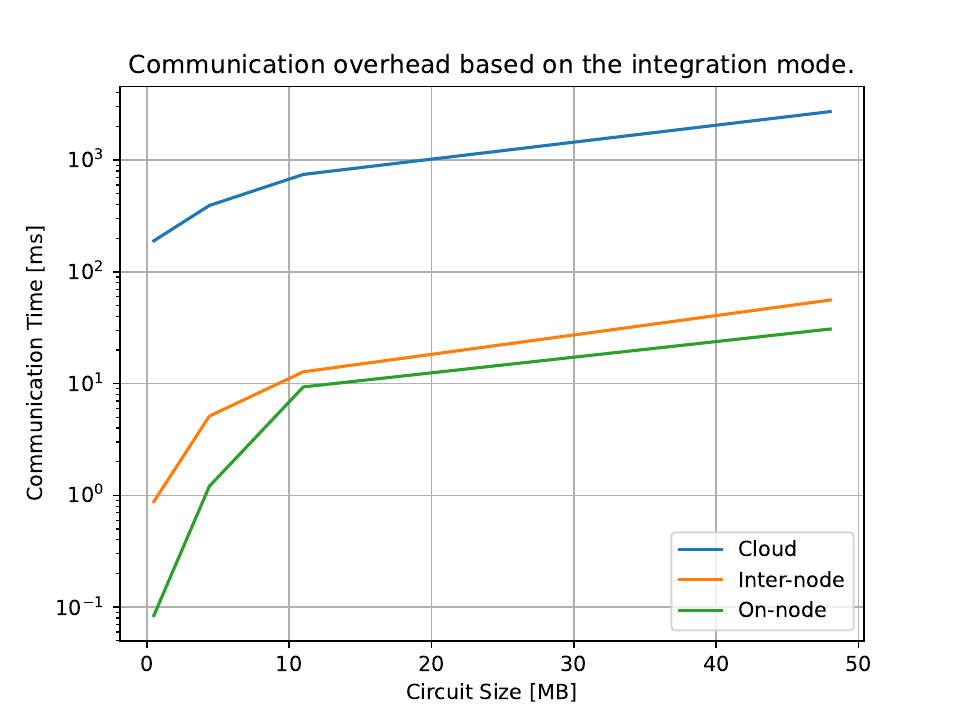}
          \caption{}
          \label{fig:qacc-comms}
        \end{subfigure}
        \begin{subfigure}{.45\textwidth}
          \centering
          \includegraphics[width=\linewidth]{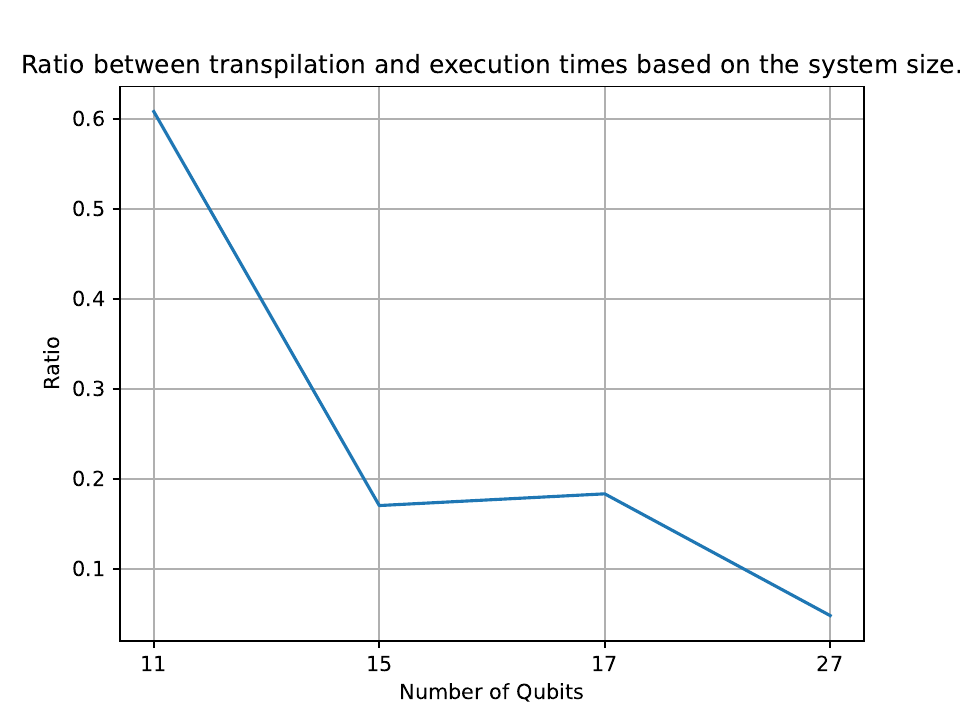}
          \caption{}
          \label{fig:qacc-runtime-ratio}
        \end{subfigure}
        \caption{(\ref{fig:qacc-comms}): Communication overhead scaling with respect to the circuit size and mode of integration. (\ref{fig:qacc-runtime-ratio}): Ratio of the transpilation to the execution times depending on the system size.}
    \label{fig:test}
    \end{figure}

    In Table \ref{tab:results-qacc-comms} and Figure \ref{fig:qacc-comms},
    the communication overheads of all three modes of integration are
    compared against each other.
    Indeed, for the small circuit sizes, the difference in the overheads
    is dramatic and on the level of orders of magnitude. As we grow the circuits,
    the difference between intra-node and inter-node slowly diminishes.
    Because the protocols used for communication over the internet network,
    large messages are split into small packets, and as such, for the circuits
    considered here, the bandwidth and latency under the load are roughly
    constant. Nevertheless, the communication overhead of the cloud integration
    is substantial and a major bottleneck for the hybrid classical-quantum
    algorithms.

    \begin{table}[t!]
        \begin{center}
            \begin{tabular}[c]{l|r|r|r}
                \hline
                \multicolumn{1}{c|}{\textbf{Molecule}} & 
                \multicolumn{1}{c|}{\textbf{Qubits}} & 
                \multicolumn{1}{c|}{\textbf{Transpilation Time [s]}} &
                \multicolumn{1}{c|}{\textbf{Quantum Runtime [s]}} \\

                \hline
                $LiH$        & 11 &  3.858    & 6.346             \\
                $NH_3$       & 15 &  28.010    & 166.353            \\
                $CH_4$       & 17 &  71.771   & 391.154            \\
                $H_2O$ dimer & 27 &  306.703  & 6352.872           \\
 
                \hline

            \end{tabular}
        \end{center}
        \caption{Transpilation time and estimated quantum runtime with respect
                to the quantum circuit and studied molecule. Quantum runtime
                is estimated by \emph{QACC-Perf} and gates characteristics outlined
                in table \ref{tab:seeqc-gates}.}
        \label{tab:results-qacc-runtime}
    \end{table}

    Finally, Table \ref{tab:results-qacc-runtime} shows the estimated
    quantum runtime values from \emph{QACC-Perf} and the associated transpilation times.
    Here, the transpilation is targeting grid topology and maps the quantum circuit
    to the basis gate set shown in Table \ref{tab:seeqc-gates}. With the increased
    qubit count and circuit size, both the transpilation and execution time 
    grow significantly.

    Comparing those results to the communication overheads, we emphasise 
    that for the communication table, the time unit is milliseconds, whereas for the
    quantum runtime, it is seconds. As such, even though the communication is 
    noticeable, it is not a major bottleneck. Indeed, the choice of the studied
    algorithm explains the situation as QPE requires little communication.
    For algorithms such as Variational Quantum Eigensolver (VQE) that have a classical and quantum component,
    the back-and-forth communication would contribute to the cost significantly.
    In general, system connectivity and resulting circuit transpilation have a far greater impact on the computation
    performance, accounting for between 5\% and 60\% of the quantum runtime
    (Figure \ref{fig:qacc-runtime-ratio}).

\section{Conclusion}
In this work, we have introduced two tools for developing hybrid HPC-QC solutions.
First, the \emph{QASC} compiler allows
users to compile general OpenQASM3.0 quantum circuits into C++ code,
which emulates the quantum program via QuEST high-performance simulator.
Additionally, we have contributed with a performance modelling tool \emph{QACC-perf}
which estimates communication costs of the offloading to the quantum
device and the runtime of the quantum program, based on the user-provided
specification. The \emph{QASC} code is available under \cite{meller_qasc_2026}. \emph{QACC-perf} is available upon request.

Both tools are scalable, lightweight and reduce barrier-to-entry in the use
of quantum resources in the domain of HPC. Such tools enable further research
into the most optimal integration strategies.

The versatility and performance of the tools have been demonstrated in the context of multiscale quantum chemistry workflows integrating HPC and quantum computing algorithms. Aside from the presented case study, the tool is generalisable to other workflows, requiring only that the generated circuits conform to OpenQASM3.0 format.

For future work, we aim to extend \emph{QASC} capabilities to support
the whole of OpenQASM3.0 specification, including gate calibration and pulse-level
instructions. Additionally, supporting pragmas and annotations would enhance
the simulator's capabilities.

Capabilities of the performance modelling tool can be further enhanced with
additional simulation use-cases to use it as a plug-in for a wider 
range of workflows.
Furthermore, we aim to benchmark and validate the modelling tool against real quantum devices. 

\section{Acknowledgements}
This work was supported by InnovateUK through project 10005792 ``Quantum Enhanced Computing Platform for Pharmaceutical R\&D – QuPharma''. For the purpose of open access, the author has applied a Creative Commons Attribution (CC BY) licence to any Author Accepted Manuscript version arising.

\section{Data availability}
The \emph{QASC} compiler is available on GitHub\footnote{\url{https://github.com/stfc/QASC}}. QACC-perf is part of a ChemShell development version and will be released in the future. Molecular configurations and scripts for generating input data for experiments can be accessed as part of our companion paper \cite{thacker_hybrid_nodate}.

\printbibliography

@article{serrano_quantum_2022,
	title = {Quantum Software Components and Platforms: Overview and Quality Assessment},
	volume = {55},
	issn = {0360-0300},
	url = {https://dl.acm.org/doi/10.1145/3548679},
	doi = {10.1145/3548679},
	shorttitle = {Quantum Software Components and Platforms},
	pages = {164:1--164:31},
	number = {8},
	journaltitle = {{ACM} Comput. Surv.},
	author = {Serrano, Manuel A. and Cruz-Lemus, José A. and Perez-Castillo, Ricardo and Piattini, Mario},
	urldate = {2024-01-30},
	date = {2022-12-23},
}

@article{bluvstein_logical_2023,
	title = {Logical quantum processor based on reconfigurable atom arrays},
	rights = {2023 The Author(s)},
	issn = {1476-4687},
	url = {https://www.nature.com/articles/s41586-023-06927-3},
	doi = {10.1038/s41586-023-06927-3},
	pages = {1--8},
	journaltitle = {Nature},
	publisher = {Nature Publishing Group},
	author = {Bluvstein, Dolev and Evered, Simon J. and Geim, Alexandra A. and Li, Sophie H. and Zhou, Hengyun and Manovitz, Tom and Ebadi, Sepehr and Cain, Madelyn and Kalinowski, Marcin and Hangleiter, Dominik and Bonilla Ataides, J. Pablo and Maskara, Nishad and Cong, Iris and Gao, Xun and Sales Rodriguez, Pedro and Karolyshyn, Thomas and Semeghini, Giulia and Gullans, Michael J. and Greiner, Markus and Vuletić, Vladan and Lukin, Mikhail D.},
	urldate = {2024-01-30},
	date = {2023-12-06},
	langid = {english},
}

@article{bauer_quantum_2020,
	title = {Quantum algorithms for quantum chemistry and quantum materials science},
	volume = {120},
	issn = {0009-2665, 1520-6890},
	url = {http://arxiv.org/abs/2001.03685},
	doi = {10.1021/acs.chemrev.9b00829},
	pages = {12685--12717},
	number = {22},
	journaltitle = {Chem. Rev.},
	author = {Bauer, Bela and Bravyi, Sergey and Motta, Mario and Chan, Garnet Kin-Lic},
	urldate = {2024-01-30},
	date = {2020-11-25},
	langid = {english},
	eprinttype = {arxiv},
	eprint = {2001.03685 [cond-mat, physics:physics, physics:quant-ph]},
}

@misc{di_meglio_quantum_2023,
	title = {Quantum Computing for High-Energy Physics: State of the Art and Challenges. Summary of the {QC}4HEP Working Group},
	url = {http://arxiv.org/abs/2307.03236},
	shorttitle = {Quantum Computing for High-Energy Physics},
	number = {{arXiv}:2307.03236},
	publisher = {{arXiv}},
	author = {Di Meglio, Alberto and Jansen, Karl and Tavernelli, Ivano and Alexandrou, Constantia and Arunachalam, Srinivasan and Bauer, Christian W. and Borras, Kerstin and Carrazza, Stefano and Crippa, Arianna and Croft, Vincent and de Putter, Roland and Delgado, Andrea and Dunjko, Vedran and Egger, Daniel J. and Fernandez-Combarro, Elias and Fuchs, Elina and Funcke, Lena and Gonzalez-Cuadra, Daniel and Grossi, Michele and Halimeh, Jad C. and Holmes, Zoe and Kuhn, Stefan and Lacroix, Denis and Lewis, Randy and Lucchesi, Donatella and Martinez, Miriam Lucio and Meloni, Federico and Mezzacapo, Antonio and Montangero, Simone and Nagano, Lento and Radescu, Voica and Ortega, Enrique Rico and Roggero, Alessandro and Schuhmacher, Julian and Seixas, Joao and Silvi, Pietro and Spentzouris, Panagiotis and Tacchino, Francesco and Temme, Kristan and Terashi, Koji and Tura, Jordi and Tuysuz, Cenk and Vallecorsa, Sofia and Wiese, Uwe-Jens and Yoo, Shinjae and Zhang, Jinglei},
	urldate = {2024-01-30},
	date = {2023-07-06},
	langid = {english},
	eprinttype = {arxiv},
	eprint = {2307.03236 [hep-ex, physics:hep-lat, physics:hep-th, physics:quant-ph]},
}

@article{preskill_quantum_2018,
	title = {Quantum Computing in the {NISQ} era and beyond},
	volume = {2},
	issn = {2521-327X},
	url = {http://arxiv.org/abs/1801.00862},
	doi = {10.22331/q-2018-08-06-79},
	pages = {79},
	journaltitle = {Quantum},
	author = {Preskill, John},
	urldate = {2024-02-06},
	date = {2018-08-06},
	langid = {english},
	eprinttype = {arxiv},
	eprint = {1801.00862 [cond-mat, physics:quant-ph]},
	 
}

@software{aleksandrowicz_qiskit_2019,
	title = {Qiskit: An Open-source Framework for Quantum Computing},
	url = {https://zenodo.org/records/2562111},
	doi = {10.5281/zenodo.2562111},
	shorttitle = {Qiskit},
	version = {0.7.2},
	publisher = {Zenodo},
	author = {Aleksandrowicz, Gadi and Alexander, Thomas and Barkoutsos, Panagiotis and Bello, Luciano and Ben-Haim, Yael and Bucher, David and Cabrera-Hernández, Francisco Jose and Carballo-Franquis, Jorge and Chen, Adrian and Chen, Chun-Fu and Chow, Jerry M. and Córcoles-Gonzales, Antonio D. and Cross, Abigail J. and Cross, Andrew and Cruz-Benito, Juan and Culver, Chris and González, Salvador De La Puente and Torre, Enrique De La and Ding, Delton and Dumitrescu, Eugene and Duran, Ivan and Eendebak, Pieter and Everitt, Mark and Sertage, Ismael Faro and Frisch, Albert and Fuhrer, Andreas and Gambetta, Jay and Gago, Borja Godoy and Gomez-Mosquera, Juan and Greenberg, Donny and Hamamura, Ikko and Havlicek, Vojtech and Hellmers, Joe and Herok, Łukasz and Horii, Hiroshi and Hu, Shaohan and Imamichi, Takashi and Itoko, Toshinari and Javadi-Abhari, Ali and Kanazawa, Naoki and Karazeev, Anton and Krsulich, Kevin and Liu, Peng and Luh, Yang and Maeng, Yunho and Marques, Manoel and Martín-Fernández, Francisco Jose and {McClure}, Douglas T. and {McKay}, David and Meesala, Srujan and Mezzacapo, Antonio and Moll, Nikolaj and Rodríguez, Diego Moreda and Nannicini, Giacomo and Nation, Paul and Ollitrault, Pauline and O'Riordan, Lee James and Paik, Hanhee and Pérez, Jesús and Phan, Anna and Pistoia, Marco and Prutyanov, Viktor and Reuter, Max and Rice, Julia and Davila, Abdón Rodríguez and Rudy, Raymond Harry Putra and Ryu, Mingi and Sathaye, Ninad and Schnabel, Chris and Schoute, Eddie and Setia, Kanav and Shi, Yunong and Silva, Adenilton and Siraichi, Yukio and Sivarajah, Seyon and Smolin, John A. and Soeken, Mathias and Takahashi, Hitomi and Tavernelli, Ivano and Taylor, Charles and Taylour, Pete and Trabing, Kenso and Treinish, Matthew and Turner, Wes and Vogt-Lee, Desiree and Vuillot, Christophe and Wildstrom, Jonathan A. and Wilson, Jessica and Winston, Erick and Wood, Christopher and Wood, Stephen and Wörner, Stefan and Akhalwaya, Ismail Yunus and Zoufal, Christa},
	urldate = {2024-03-22},
	date = {2019-01-23},
	 
}

@article{cross_openqasm_2022,
	title = {{OpenQASM} 3: A broader and deeper quantum assembly language},
	volume = {3},
	issn = {2643-6809, 2643-6817},
	url = {http://arxiv.org/abs/2104.14722},
	doi = {10.1145/3505636},
	shorttitle = {{OpenQASM} 3},
	pages = {1--50},
	number = {3},
	journaltitle = {{ACM} Transactions on Quantum Computing},
	author = {Cross, Andrew W. and Javadi-Abhari, Ali and Alexander, Thomas and de Beaudrap, Niel and Bishop, Lev S. and Heidel, Steven and Ryan, Colm A. and Sivarajah, Prasahnt and Smolin, John and Gambetta, Jay M. and Johnson, Blake R.},
	urldate = {2024-05-08},
	date = {2022-09-30},
	eprinttype = {arxiv},
	eprint = {2104.14722 [quant-ph]},
	 
}

@article{bruzewicz_trapped-ion_2019,
	title = {Trapped-Ion Quantum Computing: Progress and Challenges},
	volume = {6},
	issn = {1931-9401},
	url = {http://arxiv.org/abs/1904.04178},
	doi = {10.1063/1.5088164},
	shorttitle = {Trapped-Ion Quantum Computing},
	pages = {021314},
	number = {2},
	journaltitle = {Applied Physics Reviews},
	author = {Bruzewicz, Colin D. and Chiaverini, John and {McConnell}, Robert and Sage, Jeremy M.},
	urldate = {2024-08-07},
	date = {2019-06-01},
	eprinttype = {arxiv},
	eprint = {1904.04178 [physics, physics:quant-ph]},
	 
}

@misc{cross_open_2017,
	title = {Open Quantum Assembly Language},
	url = {http://arxiv.org/abs/1707.03429},
	number = {{arXiv}:1707.03429},
	publisher = {{arXiv}},
	author = {Cross, Andrew W. and Bishop, Lev S. and Smolin, John A. and Gambetta, Jay M.},
	urldate = {2024-08-07},
	date = {2017-07-13},
	langid = {english},
	eprinttype = {arxiv},
	eprint = {1707.03429 [quant-ph]},
	 
}

@article{humble_quantum_2021-1,
	title = {Quantum Computers for High-Performance Computing},
	volume = {41},
	issn = {1937-4143},
	url = {https://ieeexplore.ieee.org/abstract/document/9537178},
	doi = {10.1109/MM.2021.3099140},
	pages = {15--23},
	number = {5},
	journaltitle = {{IEEE} Micro},
	author = {Humble, Travis S. and {McCaskey}, Alexander and Lyakh, Dmitry I. and Gowrishankar, Meenambika and Frisch, Albert and Monz, Thomas},
	urldate = {2024-01-30},
	date = {2021-09},
	 
}

@article{arute_quantum_2019,
	title = {Quantum supremacy using a programmable superconducting processor},
	volume = {574},
	rights = {2019 The Author(s), under exclusive licence to Springer Nature Limited},
	issn = {1476-4687},
	url = {https://www.nature.com/articles/s41586-019-1666-5},
	doi = {10.1038/s41586-019-1666-5},
	pages = {505--510},
	number = {7779},
	journaltitle = {Nature},
	publisher = {Nature Publishing Group},
	author = {Arute, Frank and Arya, Kunal and Babbush, Ryan and Bacon, Dave and Bardin, Joseph C. and Barends, Rami and Biswas, Rupak and Boixo, Sergio and Brandao, Fernando G. S. L. and Buell, David A. and Burkett, Brian and Chen, Yu and Chen, Zijun and Chiaro, Ben and Collins, Roberto and Courtney, William and Dunsworth, Andrew and Farhi, Edward and Foxen, Brooks and Fowler, Austin and Gidney, Craig and Giustina, Marissa and Graff, Rob and Guerin, Keith and Habegger, Steve and Harrigan, Matthew P. and Hartmann, Michael J. and Ho, Alan and Hoffmann, Markus and Huang, Trent and Humble, Travis S. and Isakov, Sergei V. and Jeffrey, Evan and Jiang, Zhang and Kafri, Dvir and Kechedzhi, Kostyantyn and Kelly, Julian and Klimov, Paul V. and Knysh, Sergey and Korotkov, Alexander and Kostritsa, Fedor and Landhuis, David and Lindmark, Mike and Lucero, Erik and Lyakh, Dmitry and Mandrà, Salvatore and {McClean}, Jarrod R. and {McEwen}, Matthew and Megrant, Anthony and Mi, Xiao and Michielsen, Kristel and Mohseni, Masoud and Mutus, Josh and Naaman, Ofer and Neeley, Matthew and Neill, Charles and Niu, Murphy Yuezhen and Ostby, Eric and Petukhov, Andre and Platt, John C. and Quintana, Chris and Rieffel, Eleanor G. and Roushan, Pedram and Rubin, Nicholas C. and Sank, Daniel and Satzinger, Kevin J. and Smelyanskiy, Vadim and Sung, Kevin J. and Trevithick, Matthew D. and Vainsencher, Amit and Villalonga, Benjamin and White, Theodore and Yao, Z. Jamie and Yeh, Ping and Zalcman, Adam and Neven, Hartmut and Martinis, John M.},
	urldate = {2024-08-12},
	date = {2019-10},
	langid = {english},
	 
}

@article{alexeev_quantum-centric_2024,
	title = {Quantum-centric supercomputing for materials science: A perspective on challenges and future directions},
	volume = {160},
	issn = {0167-739X},
	url = {https://www.sciencedirect.com/science/article/pii/S0167739X24002012},
	doi = {10.1016/j.future.2024.04.060},
	shorttitle = {Quantum-centric supercomputing for materials science},
	pages = {666--710},
	journaltitle = {Future Generation Computer Systems},
	author = {Alexeev, Yuri and Amsler, Maximilian and Barroca, Marco Antonio and Bassini, Sanzio and Battelle, Torey and Camps, Daan and Casanova, David and Choi, Young Jay and Chong, Frederic T. and Chung, Charles and Codella, Christopher and Córcoles, Antonio D. and Cruise, James and Di Meglio, Alberto and Duran, Ivan and Eckl, Thomas and Economou, Sophia and Eidenbenz, Stephan and Elmegreen, Bruce and Fare, Clyde and Faro, Ismael and Fernández, Cristina Sanz and Ferreira, Rodrigo Neumann Barros and Fuji, Keisuke and Fuller, Bryce and Gagliardi, Laura and Galli, Giulia and Glick, Jennifer R. and Gobbi, Isacco and Gokhale, Pranav and de la Puente Gonzalez, Salvador and Greiner, Johannes and Gropp, Bill and Grossi, Michele and Gull, Emanuel and Healy, Burns and Hermes, Matthew R. and Huang, Benchen and Humble, Travis S. and Ito, Nobuyasu and Izmaylov, Artur F. and Javadi-Abhari, Ali and Jennewein, Douglas and Jha, Shantenu and Jiang, Liang and Jones, Barbara and de Jong, Wibe Albert and Jurcevic, Petar and Kirby, William and Kister, Stefan and Kitagawa, Masahiro and Klassen, Joel and Klymko, Katherine and Koh, Kwangwon and Kondo, Masaaki and Kürkçüog̃lu, Dog̃a Murat and Kurowski, Krzysztof and Laino, Teodoro and Landfield, Ryan and Leininger, Matt and Leyton-Ortega, Vicente and Li, Ang and Lin, Meifeng and Liu, Junyu and Lorente, Nicolas and Luckow, Andre and Martiel, Simon and Martin-Fernandez, Francisco and Martonosi, Margaret and Marvinney, Claire and Medina, Arcesio Castaneda and Merten, Dirk and Mezzacapo, Antonio and Michielsen, Kristel and Mitra, Abhishek and Mittal, Tushar and Moon, Kyungsun and Moore, Joel and Mostame, Sarah and Motta, Mario and Na, Young-Hye and Nam, Yunseong and Narang, Prineha and Ohnishi, Yu-ya and Ottaviani, Daniele and Otten, Matthew and Pakin, Scott and Pascuzzi, Vincent R. and Pednault, Edwin and Piontek, Tomasz and Pitera, Jed and Rall, Patrick and Ravi, Gokul Subramanian and Robertson, Niall and Rossi, Matteo A. C. and Rydlichowski, Piotr and Ryu, Hoon and Samsonidze, Georgy and Sato, Mitsuhisa and Saurabh, Nishant and Sharma, Vidushi and Sharma, Kunal and Shin, Soyoung and Slessman, George and Steiner, Mathias and Sitdikov, Iskandar and Suh, In-Saeng and Switzer, Eric D. and Tang, Wei and Thompson, Joel and Todo, Synge and Tran, Minh C. and Trenev, Dimitar and Trott, Christian and Tseng, Huan-Hsin and Tubman, Norm M. and Tureci, Esin and Valiñas, David García and Vallecorsa, Sofia and Wever, Christopher and Wojciechowski, Konrad and Wu, Xiaodi and Yoo, Shinjae and Yoshioka, Nobuyuki and Yu, Victor Wen-zhe and Yunoki, Seiji and Zhuk, Sergiy and Zubarev, Dmitry},
	urldate = {2024-08-15},
	date = {2024-11-01},
	 
}

@misc{bergholm_pennylane_2022,
	title = {{PennyLane}: Automatic differentiation of hybrid quantum-classical computations},
	url = {http://arxiv.org/abs/1811.04968},
	doi = {10.48550/arXiv.1811.04968},
	shorttitle = {{PennyLane}},
	number = {{arXiv}:1811.04968},
	publisher = {{arXiv}},
	author = {Bergholm, Ville and Izaac, Josh and Schuld, Maria and Gogolin, Christian and Ahmed, Shahnawaz and Ajith, Vishnu and Alam, M. Sohaib and Alonso-Linaje, Guillermo and {AkashNarayanan}, B. and Asadi, Ali and Arrazola, Juan Miguel and Azad, Utkarsh and Banning, Sam and Blank, Carsten and Bromley, Thomas R. and Cordier, Benjamin A. and Ceroni, Jack and Delgado, Alain and Matteo, Olivia Di and Dusko, Amintor and Garg, Tanya and Guala, Diego and Hayes, Anthony and Hill, Ryan and Ijaz, Aroosa and Isacsson, Theodor and Ittah, David and Jahangiri, Soran and Jain, Prateek and Jiang, Edward and Khandelwal, Ankit and Kottmann, Korbinian and Lang, Robert A. and Lee, Christina and Loke, Thomas and Lowe, Angus and {McKiernan}, Keri and Meyer, Johannes Jakob and Montañez-Barrera, J. A. and Moyard, Romain and Niu, Zeyue and O'Riordan, Lee James and Oud, Steven and Panigrahi, Ashish and Park, Chae-Yeun and Polatajko, Daniel and Quesada, Nicolás and Roberts, Chase and Sá, Nahum and Schoch, Isidor and Shi, Borun and Shu, Shuli and Sim, Sukin and Singh, Arshpreet and Strandberg, Ingrid and Soni, Jay and Száva, Antal and Thabet, Slimane and Vargas-Hernández, Rodrigo A. and Vincent, Trevor and Vitucci, Nicola and Weber, Maurice and Wierichs, David and Wiersema, Roeland and Willmann, Moritz and Wong, Vincent and Zhang, Shaoming and Killoran, Nathan},
	urldate = {2024-10-28},
	date = {2022-07-29},
	eprinttype = {arxiv},
	eprint = {1811.04968},
	 
}

@inproceedings{kaya_software_2024,
	location = {Atlanta {GA} {USA}},
	title = {A Software Platform to Support Disaggregated Quantum Accelerators},
	url = {https://conferences.computer.org/sc-wpub/pdfs/SC-W2024-6oZmigAQfgJ1GhPL0yE3pS/555400b646/555400b646.pdf},
	doi = {10.1109/SCW63240.2024.00205},
	eventtitle = {{SC}-W 2024: Workshops of The International Conference on High Performance Computing, Network, Storage, and Analysis},
	booktitle = {Workshop: International Workshop on {RESource} {DISaggregation} in High Performance Computing ({RESDIS})},
	publisher = {{IEEE}},
	author = {Kaya, Ercüment},
	date = {2024-11-17},
	langid = {english},
	 
}

@book{nielsen_quantum_2010,
	title = {Quantum Computation and Quantum Information: 10th Anniversary Edition},
	isbn = {978-0-511-97666-7},
	url = {https://www.cambridge.org/highereducation/books/quantum-computation-and-quantum-information/01E10196D0A682A6AEFFEA52D53BE9AE},
	doi = {10.1017/CBO9780511976667},
	shorttitle = {Quantum Computation and Quantum Information},
	publisher = {Cambridge University Press},
	author = {Nielsen, Michael A. and Chuang, Isaac L.},
	urldate = {2025-11-20},
	date = {2010-12-09},
	langid = {english},
	 
}

@article{culler_logp_1996,
	title = {{LogP}: a practical model of parallel computation},
	volume = {39},
	issn = {0001-0782, 1557-7317},
	url = {https://dl.acm.org/doi/10.1145/240455.240477},
	doi = {10.1145/240455.240477},
	shorttitle = {{LogP}},
	pages = {78--85},
	number = {11},
	journaltitle = {Commun. {ACM}},
	author = {Culler, David E. and Karp, Richard M. and Patterson, David and Sahay, Abhijit and Santos, Eunice E. and Schauser, Klaus Erik and Subramonian, Ramesh and Von Eicken, Thorsten},
	urldate = {2025-11-20},
	date = {1996-11},
	langid = {english},
	 
}

@inproceedings{ino_loggps_2001,
	location = {New York, {NY}, {USA}},
	title = {{LogGPS}: a parallel computational model for synchronization analysis},
	isbn = {978-1-58113-346-2},
	url = {https://dl.acm.org/doi/10.1145/379539.379592},
	doi = {10.1145/379539.379592},
	series = {{PPoPP} '01},
	shorttitle = {{LogGPS}},
	pages = {133--142},
	booktitle = {Proceedings of the eighth {ACM} {SIGPLAN} symposium on Principles and practices of parallel programming},
	publisher = {Association for Computing Machinery},
	author = {Ino, Fumihiko and Fujimoto, Noriyuki and Hagihara, Kenichi},
	urldate = {2025-11-20},
	date = {2001-06-18},
	 
}

@report{williams_roofline_2009,
	title = {Roofline: An Insightful Visual Performance Model for Floating-Point Programs and Multicore Architectures},
	url = {http://www.osti.gov/servlets/purl/1407078/},
	doi = {10.2172/1407078},
	shorttitle = {Roofline},
	pages = {1407078},
	number = {1407078},
	author = {Williams, Samuel and Waterman, Andrew and Patterson, David},
	urldate = {2025-11-20},
	date = {2009-09-01},
	langid = {english},
	 
}

@article{jones_quest_2019,
	title = {{QuEST} and High Performance Simulation of Quantum Computers},
	volume = {9},
	rights = {2019 The Author(s)},
	issn = {2045-2322},
	url = {https://www.nature.com/articles/s41598-019-47174-9},
	doi = {10.1038/s41598-019-47174-9},
	pages = {10736},
	number = {1},
	journaltitle = {Sci Rep},
	publisher = {Nature Publishing Group},
	author = {Jones, Tyson and Brown, Anna and Bush, Ian and Benjamin, Simon C.},
	urldate = {2025-11-20},
	date = {2019-07-24},
	langid = {english},
	 
}

@article{lu_open-source_2019,
	title = {Open-Source, Python-Based Redevelopment of the {ChemShell} Multiscale {QM}/{MM} Environment},
	volume = {15},
	rights = {All rights reserved},
	issn = {1549-9618},
	url = {https://doi.org/10.1021/acs.jctc.8b01036},
	doi = {10.1021/acs.jctc.8b01036},
	pages = {1317--1328},
	number = {2},
	journaltitle = {J. Chem. Theory Comput.},
	publisher = {American Chemical Society},
	author = {Lu, You and Farrow, Matthew R. and Fayon, Pierre and Logsdail, Andrew J. and Sokol, Alexey A. and Catlow, C. Richard A. and Sherwood, Paul and Keal, Thomas W.},
	urldate = {2022-10-31},
	date = {2019-02-12},
}

@article{somma_quantum_2019,
	title = {Quantum eigenvalue estimation via time series analysis},
	volume = {21},
	issn = {1367-2630},
	url = {https://dx.doi.org/10.1088/1367-2630/ab5c60},
	doi = {10.1088/1367-2630/ab5c60},
	pages = {123025},
	number = {12},
	journaltitle = {New J. Phys.},
	publisher = {{IOP} Publishing},
	author = {Somma, Rolando D.},
	urldate = {2024-02-06},
	date = {2019-12},
	langid = {english},
}

@article{buchs_role_2026,
	title = {The role of quantum computing in advancing scientific high-performance computing: A perspective from the {ADAC} institute},
	volume = {182},
	issn = {0167-739X},
	url = {https://www.sciencedirect.com/science/article/pii/S0167739X26001214},
	doi = {10.1016/j.future.2026.108487},
	shorttitle = {The role of quantum computing in advancing scientific high-performance computing},
	pages = {108487},
	journaltitle = {Future Generation Computer Systems},
	author = {Buchs, Gilles and Beck, Thomas L. and Bennink, Ryan S. and Claudino, Daniel and Delgado, Andrea and Fadel, Nur Aiman and Groszkowski, Peter and Hamilton, Kathleen E. and Humble, Travis S. and Kumar, Neeraj and Li, Ang and Lotshaw, Phillip C. and Mukkula, Olli and Takano, Ryousei and Saxena, Amit and Suh, In-Saeng and Tsuji, Miwako and Van Beeumen, Roel and Varetto, Ugo and Wang, Yan and Yamazaki, Kazuya and Johansson, Mikael P.},
	urldate = {2026-04-27},
	date = {2026-09-01},
	 
}

@article{perez_castillo_software_2021,
	title = {Software modernization to embrace quantum technology},
	volume = {151},
	issn = {0965-9978},
	url = {https://www.sciencedirect.com/science/article/pii/S0965997820309790},
	doi = {10.1016/j.advengsoft.2020.102933},
	pages = {102933},
	journaltitle = {Advances in Engineering Software},
	author = {Pérez-Castillo, Ricardo and Serrano, Manuel A. and Piattini, Mario},
	urldate = {2026-04-27},
	date = {2021-01-01},
	 
}

@article{elsharkawy_integration_2025,
	title = {Integration of Quantum Accelerators with High Performance Computing—A Review of Quantum Programming Tools},
	volume = {6},
	url = {https://dl.acm.org/doi/10.1145/3743149},
	doi = {10.1145/3743149},
	pages = {24:1--24:46},
	number = {3},
	journaltitle = {{ACM} Transactions on Quantum Computing},
	author = {Elsharkawy, Amr and To, Xiao-Ting Michelle and Seitz, Philipp and Chen, Yanbin and Stade, Yannick and Geiger, Manuel and Huang, Qunsheng and Guo, Xiaorang and Ansari, Muhammad Arslan and Mendl, Christian B. and Kranzlmüller, Dieter and Schulz, Martin},
	urldate = {2026-04-27},
	date = {2025-07-12},
	 
}

@article{sun_recent_2020,
	title = {Recent developments in the {PySCF} program package},
	volume = {153},
	issn = {0021-9606},
	url = {https://doi.org/10.1063/5.0006074},
	doi = {10.1063/5.0006074},
	pages = {024109},
	number = {2},
	journaltitle = {J. Chem. Phys.},
	author = {Sun, Qiming and Zhang, Xing and Banerjee, Samragni and Bao, Peng and Barbry, Marc and Blunt, Nick S. and Bogdanov, Nikolay A. and Booth, George H. and Chen, Jia and Cui, Zhi-Hao and Eriksen, Janus J. and Gao, Yang and Guo, Sheng and Hermann, Jan and Hermes, Matthew R. and Koh, Kevin and Koval, Peter and Lehtola, Susi and Li, Zhendong and Liu, Junzi and Mardirossian, Narbe and {McClain}, James D. and Motta, Mario and Mussard, Bastien and Pham, Hung Q. and Pulkin, Artem and Purwanto, Wirawan and Robinson, Paul J. and Ronca, Enrico and Sayfutyarova, Elvira R. and Scheurer, Maximilian and Schurkus, Henry F. and Smith, James E. T. and Sun, Chong and Sun, Shi-Ning and Upadhyay, Shiv and Wagner, Lucas K. and Wang, Xiao and White, Alec and Whitfield, James Daniel and Williamson, Mark J. and Wouters, Sebastian and Yang, Jun and Yu, Jason M. and Zhu, Tianyu and Berkelbach, Timothy C. and Sharma, Sandeep and Sokolov, Alexander Yu. and Chan, Garnet Kin-Lic},
	urldate = {2026-01-05},
	date = {2020-07-09},
}

@article{thacker_hybrid_nodate,
    author = {Thacker, Joseph C. R. and Meller, Mateusz and Popescu, Alberto and Patterson, Andrew and Izsák, Róbert and Szeremi, Vendel and Keal, Thomas W.},
    title = {Hybrid HPC and
Quantum Computing Framework for Multiscale
Modeling of Chemical Systems},
    journal = {Journal of Chemical Theory and Computation},
    volume = {22},
    number = {16},
    pages = {8201-8214},
    year = {2026},
    month = {08},
    issn = {1549-9618},
    doi = {10.1021/acs.jctc.6c00542},
    url = {https://doi.org/10.1021/acs.jctc.6c00542},
    eprint = {https://pubs.acs.org/jctcce/article-pdf/22/16/8201/66406147/acs.jctc.6c00542.pdf},
}

@article{jordan_quantum_2026,
	title = {Quantum Computer Controlled by Superconducting Digital Electronics at Millikelvin Temperature},
	volume = {9},
	issn = {2520-1131},
	url = {http://arxiv.org/abs/2503.09879},
	doi = {10.1038/s41928-026-01576-6},
	pages = {287--294},
	number = {3},
	journaltitle = {Nat Electron},
	author = {Jordan, Caleb and Bernhardt, Jacob and Rahamim, Joseph and Kirichenko, Alex and Bharadwaj, Karthik and Fry-Bouriaux, Louis and Somoroff, Aaron and Porsch, Katie and Tsai, Kan-Ting and Walter, Jason and Weis, Adam and Yu, Meng-Ju and Renzullo, Mario and Javelle, Jerome and Checkley, Chris and Mukhanov, Oleg and Yohannes, Daniel and Vernik, Igor and Han, Shu-Jen},
	urldate = {2026-05-05},
	date = {2026-03-10},
	eprinttype = {arxiv},
	eprint = {2503.09879 [quant-ph]},
	 
}

@report{johansson_quantum_2021,
	title = {Quantum Computing – A European Perspective},
	url = {https://zenodo.org/records/5547408},
	doi = {10.5281/zenodo.5547408},
	institution = {Zenodo},
	author = {Johansson, Mikael P. and Krishnasamy, Ezhilmathi and Meyer, Norbert and Piechurski, Christelle},
	urldate = {2026-05-05},
	date = {2021-09-30},
	 
}

@book{parr_definitive_2013,
	edition = {2nd},
	title = {The Definitive {ANTLR} 4 Reference},
	isbn = {1934356999},
	pagetotal = {328},
	publisher = {Pragmatic Bookshelf},
	author = {Parr, Terence},
	date = {2013},
}

@software{breuer_jupiter_2024,
	title = {{JUPITER} Benchmark Suite: {OSU} Micro-Benchmarks},
	url = {https://zenodo.org/records/12788752},
	doi = {10.5281/zenodo.12788752},
	shorttitle = {{JUPITER} Benchmark Suite},
	publisher = {Zenodo},
	author = {Breuer, Thomas and Herten, Andreas and Achilles, Sebastian},
	urldate = {2026-05-05},
	date = {2024-07-20},
}

@software{meller_qasc_2026,
	title = {{QASC} Quantum Assembly Compiler},
	url = {https://github.com/stfc/QASC},
    doi = {10.5281/zenodo.23065738},
	shorttitle = {{QASC}},
	publisher = {Zenodo},
	author = {Meller, Mateusz and Thacker, Joseph C. R. and Keal, Thomas W. and Szeremi, Vendel},
	urldate = {2026-08-20},
	date = {2026-08-20},
}

\end{document}